\documentclass[twocolumn,showpacs,prb]{revtex4}
\usepackage{graphicx}

\begin{document}

\preprint{}
\title{
Field dependence of the vortex structure in chiral $p$-wave superconductors
}

\author{Masanori Ichioka} 
\email{oka@mp.okayama-u.ac.jp}
\author{Kazushige Machida}
\email{machida@mp.okayama-u.ac.jp}
\affiliation{Department of Physics, Okayama University,
         Okayama 700-8530, Japan}

\date{February 28, 2002
}

\begin{abstract}
To investigate the different  vortex structure between two chiral pairing 
$p_x \pm{\rm i}p_y$,  we calculate the pair potential, 
the internal field, the local density of states, and free energy 
in the vortex lattice state based on the quasiclassical Eilenberger 
theory, and analyze the magnetic field dependence. 
The induced opposite chiral component of the pair potential plays an  
important role in the vortex structure. 
It also produces $\sqrt{H}$-behavior of the zero-energy density of states 
at higher field. 
These results are helpful when we understand the vortex states in 
${\rm Sr_2RuO_4}$. 
\end{abstract}



\maketitle 
\section{Introduction}
\label{sec:introduction}

For the superconducting state in 
quasi-two-dimensional metal ${\rm Sr_2RuO_4}$,
the pairing symmetry is suggested to be the chiral $p$-wave pairing 
with the basic form $\Delta_{\pm}\sim p_x \pm{\rm i} p_y$ and 
inplane equal-spin pairing.~\cite{Maeno,Rice}
For the experimental evidence, 
the spin triplet pairing is supported by 
the ${\rm ^{17}O}$-NMR measurement, which reported that there is 
no reduction of the Knight shift in the superconducting 
state.~\cite{Ishida} 
The pairing state with broken time reversal symmetry is claimed 
by the $\mu$SR measurement, which reported that spontaneous magnetic 
moment is induced in the superconducting state.~\cite{Luke}  

Since the $\Delta_{+}$ state and the $\Delta_{-}$ state are degenerate 
at zero field, we expect the domain formation of the two states 
$\Delta_{+}$ and $\Delta_{-}$. 
We call them as the $p_+$-wave domain and the $p_-$-wave domain, respectively. 
This degeneracy is lifted under external magnetic field perpendicular 
to the basal plane, since $\Delta_{\pm}$ is the broken time reversal 
symmetry state with an orbital angular momentum along the $z$-axis.  
Then, the vortex in the mixed state 
shows the different structure for the $p_+$-wave and the $p_-$-wave domains. 
We consider the case when 
${\bf H} \parallel \hat{z}$ for $e>0$ ($2e$ is the charge of the Cooper 
pair) or equivalently when ${\bf H} \parallel -\hat{z}$ for $e<0$. 
In these cases, there appear winding $+1$ vortices. 
It is parallel (antiparallel) to the internal winding of the 
Cooper pair in the $p_+$-wave ($p_-$-wave) domain. 
The information of the vortex structure for the $p_\pm$-wave domain 
is important to analyze the chirality $p_\pm$ of each domain. 

The differences of the vortex structure for the $p_+$-wave and the 
$p_-$-wave domains have been studied by the two component Ginzburg-Laudau 
(GL) theory,~\cite{Heeb,HeebD} 
the quasiclassical theory,~\cite{Kato,KatoHayashi}  
and the Bogoliubov-de Gennes (BdG) 
theory.~\cite{MatsumotoHeeb,HeebD,Matsumoto,TakigawaP}  
In the chiral $p$-wave superconductors, it is important to consider 
the opposite chiral component 
$\Delta_{\mp}$ which is induced around the vortex of the dominant 
$\Delta_{\pm}$ component in the $p_\pm$-wave domains. 
This induced component shows different spatial structure for 
the $p_+$-wave and the $p_-$-wave domains. 
This is the origin of the different vortex structure. 
There is a free energy difference between the $p_+$-wave and 
the $p_-$-wave domain cases in the vortex state, 
which leads to the different upper critical field $H_{c2}$ 
for these chiral states. 
The estimation of $H_{c2}$ was reported by Scharnberg and Klemm 
in the isotropic three dimensional Fermi surface 
case.\cite{Scharnberg,ScharnbergL}
Following their results, 
$H_{c2}$ of the $p_+$-wave domain, which is noted as ABM state, 
is near that of the isotropic $s$-wave pairing case, 
because the opposite $\Delta_{-}$-component is not induced 
along $H_{c2}$-line. 
On the other hand, the vortex state in the $p_-$-wave domain, 
which is noted as the generalized ABM state or the SK state, 
gives higher $H_{c2}$ because 
there is large induced $\Delta_{+}$-component. 
That is, we can say that the superconductivity survives until high 
field in the $p_-$-wave domain case by the enhancement due to 
the induced opposite chiral component. 
Since the difference of $H_{c2}$ is large (about twice) for these 
domain cases, 
it is important to know the $H$ (external magnetic field) dependence 
of the vortex structure for $\Delta_{+}$ and $\Delta_{-}$  continuously 
for all $H$ regions, in order to study the chiral-dependent properties. 

In this paper, we investigate the field dependence of the vortex 
structure for the $p_\pm$-wave domain cases, based on the 
quasiclassical Eilenberger theory.\cite{Eilenberger} 
Our calculation method for the vortex lattice state was established 
for the study of the $d_{x^2-y^2}$-wave pairing case in high-$T_c$ 
superconductors.\cite{IchiokaQCLd1,IchiokaQCLd2,IchiokaJS,
IchiokaQCLs,IchiokaD,IchiokaInd} 
We can calculate both the pair potential and the vector potential 
selfconsistently.  
In this paper, we consider the case of small GL parameter 
$\kappa=2.7$ appropriate to ${\rm Sr_2 Ru O_4}$.\cite{Akima}  
In the quasiclassical theory, we can also consider the quasiparticle states 
around the vortex, such as the local density of states (LDOS).  
Among them, $H$-dependence of the spatially averaged 
zero energy density of states (DOS) $N(0)$ is important. 
It is accessible by the specific heat measurement. 
For example, in the $d_{x^2-y^2}$-wave pairing such as in high $T_c$ 
superconductors, there is a relation $N(0)\sim \sqrt{H}$ due to the 
line node of the superconducting gap for 
${\bf H} \parallel c$.\cite{Volovik,Moler,Fisher} 
It is because low energy quasiparticles propagating to the node 
direction can extend far from vortex. 
In the relation $N(0)\sim H^\alpha$ in the $s$-wave pairing case, 
$\alpha$ can be smaller than 1. 
It is related to the field dependence of the vortex core 
radius.\cite{Ramirez,Hedo,Golubov,SonierNbSe2,SonierR}
These behaviors could be confirmed by the calculation based on the 
quasiclassical theory.\cite{IchiokaQCLd1}
In ${\rm Sr_2 Ru O_4}$, the specific heat  measurement 
suggests the relation $N(0)\sim \sqrt{H}$ at higher field, 
giving the discussion on the possibility of the line node.\cite{Deguchi}  
Then, it is important to examine the origin of the 
$\sqrt{H}$-behavior. 

There are detailed discussions on the pairing function of the chiral 
$p$-wave for ${\rm Sr_2 Ru O_4}$, 
including the gap anisotropy and the orbital 
dependence.\cite{Machida,Sigrist,Hasegawa,Graf,Zhitomirsky,Ng,Miyake}  
But, here, we consider the fundamental case of the simple isotropic 
gap function 
$\Delta_\pm =(p_x \pm {\rm i} p_y)/ p_{\rm F}={\rm e}^{\pm {\rm i}\theta}$ 
on the two-dimensional isotropic Fermi surface 
$(p_x,p_y)=p_{\rm F}(\cos\theta,\sin\theta)$ 
in order to focus on the chirality dependence, without including the 
anisotropy effect of the superconducting gap or Fermi surfaces. 

After describing our formulation of the quasiclassical theory  
in Sec. \ref{sec:formulation}, we evaluate the free energy in 
Sec. \ref{sec:fenergy}. 
The pair potential and internal field structure of the 
vortex lattice state are studied in Sec. \ref{sec:vortex}. 
The low energy quasiparticle states are examined in  Sec. \ref{sec:LDOS} 
by considering the LDOS and the field dependence of the DOS. 
The last section is devoted to summary and discussions.

\section{quasiclassical Eilenberger theory}
\label{sec:formulation}

Our calculation is performed by extending the quasiclassical method 
for the vortex lattice state in the $d_{x^2-y^2}$-wave pairing 
case~\cite{IchiokaQCLd1,IchiokaQCLd2,IchiokaJS,
IchiokaQCLs,IchiokaD,IchiokaInd}
to the chiral $p$-wave pairing case.  
For the details of the calculation method, also see 
Refs. \onlinecite{KleinJLTP,KleinPRB,Pottinger}. 
We consider the case of the clean limit and cylindrical Fermi surface.
First, to obtain the pair potential $\Delta(\theta,{\bf r})$ and 
vector potential ${\bf A}({\bf r})$ selfconsistently, 
we solve the Eilenberger equation in the Matsubara
frequency $\omega_n=(2n+1)\pi T$ for the quasiclassical Green's functions
$g({\rm i}\omega_n,\theta,{\bf r})$, $f({\rm i}\omega_n,\theta,{\bf r})$ and
$f^\dagger({\rm i}\omega_n,\theta,{\bf r})$,
where ${\bf r}$ is the center of mass coordinate of a Cooper pair.
The direction of the relative momentum of the Cooper pair, 
$\hat{\bf k}={\bf k}/|{\bf k}|$, is denoted by an angle $\theta$
measured from the $x$ axis.
The Eilenberger equation is given by~\cite{Eilenberger}
\begin{eqnarray} &&
\Bigl\{ \omega_n +{{\rm i} \over 2} {\bf v}_{\rm F}\cdot
\Bigl(\frac{\nabla}{\rm i}+ \frac{2\pi}{\phi_0} {\bf A}({\bf r}) \Bigr) \Bigr\}
f({\rm i}\omega_n,\theta,{\bf r})
\nonumber \\ &&
= \Delta(\theta,{\bf r}) g({\rm i}\omega_n,\theta,{\bf r}) ,
\label{eq:eil-1} \\  &&
\Bigl\{ \omega_n -{{\rm i} \over 2} {\bf v}_{\rm F}\cdot
\Bigl(\frac{\nabla}{\rm i}-\frac{2\pi}{\phi_0} {\bf A}({\bf r}) \Bigr) \Bigr\}
f^\dagger({\rm i}\omega_n,\theta,{\bf r})
\nonumber \\ &&
= \Delta^\ast(\theta,{\bf r}) g({\rm i}\omega_n,\theta,{\bf r}),
\label{eq:eil-2} \\  &&
g({\rm i}\omega_n,\theta,{\bf r}) =[1- f({\rm i}\omega_n,\theta,{\bf r})
    f^\dagger({\rm i}\omega_n,\theta,{\bf r}) ]^{1/2},
\label{eq:eil-3}
\end{eqnarray}
where ${\rm Re} g({\rm i}\omega_n,\theta,{\bf r}) > 0$, 
the Fermi velocity ${\bf v}_{\rm F}=v_{\rm F}\hat{\bf k}$, 
the flux quantum $\phi_0$ and $e=-|e|$. 
In the symmetric gauge, 
${\bf A}({\bf r})=\frac{1}{2} {\bf H} \times {\bf r} + {\bf a}({\bf r})$, 
where ${\bf H}=(0,0,H)$ is a uniform field and ${\bf a}({\bf r})$ is 
related to the internal field ${\bf h}({\bf r})=(0,0,h({\bf r}))$
as ${\bf h}({\bf r})=\nabla\times {\bf a}({\bf r})$.
For the coupling to a magnetic field, we neglect the paramagnetic coupling 
to the spin, and consider the effect of the orbital coupling in the 
vector potential terms. 

The self-consistent conditions for $\Delta(\theta,{\bf r})$ and 
${\bf a}({\bf r})$ are given as
\begin{equation}
\Delta(\theta,{\bf r})= N_0
2 \pi T \sum_{\omega_n>0} \int_0^{2\pi}{{\rm d}\theta' \over 2\pi}
V(\theta',\theta)f({\rm i}\omega_n,\theta',{\bf r}) ,
\label{eq:self-d}
\end{equation}
\begin{equation}
\nabla\times\nabla\times{\bf a}({\bf r})
= - \frac{\pi\phi_0}{\kappa^2 \Delta_0 \xi_0^3} 2\pi T \sum_{\omega_n>0}
\int_0^{2\pi}{{\rm d}\theta \over 2\pi}{\hat{\bf k} \over {\rm i}}
g({\rm i}\omega_n,\theta,{\bf r}) ,
\label{eq:self-a}
\end{equation}
with  the pairing interaction $V(\theta',\theta)$ and 
$\kappa=(7 \zeta(3)/72)^{1/2}(\Delta_0/ T_c)\kappa_{\rm BCS}$
with Rieman's zeta function $\zeta(3)$.
$N_0$ is the density of states at the Fermi surface, 
$\Delta_0$ is the uniform gap at $T=0$, and 
$\kappa_{\rm BCS}$ is the GL parameter in the BCS theory.  
We set the energy cutoff $\omega_c=20 T_c$ and $\kappa_{\rm BCS}=2.7$. 
In the following, energies and lengths are measured in units of $\Delta_0$
and $\xi_0=v_{\rm F}/\Delta_0 =\pi \xi_{\rm BCS}$ 
($\xi_{\rm BCS}$ is the BCS coherence length), respectively.
The magnetic fields are measured in units of $\phi_0/\xi_0^2$. 

By solving Eqs. (\ref{eq:eil-1})-(\ref{eq:eil-3}) in the so-called explosion
method, we estimate the quasiclassical Green's functions at $41\times 41$ 
discretized points in a unit cell of the vortex lattice.
Using the symmetry relation~\cite{IchiokaQCLs,KleinJLTP} 
described in Appendix, we can reduce the range of ${\bf r}$ and $\theta$ 
in the calculation solving Eqs. (\ref{eq:eil-1})-(\ref{eq:eil-3}). 
We obtain new $\Delta(\theta,{\bf r})$ and ${\bf a}({\bf r})$ from
Eqs. (\ref{eq:self-d}) and (\ref{eq:self-a}), and use them at the next step
calculation of Eqs. (\ref{eq:eil-1})-(\ref{eq:eil-3}).
This iteration procedure is repeated until sufficiently
selfconsistent solution is obtained.
When we consider the lattice transformation 
${\bf R}=m {\bf r}_1+n {\bf r}_2$ ($m$, $n$ : integers) with the unit 
vectors ${\bf r}_1=(a_x,0)$, ${\bf r}_1=(\zeta a_x,a_y)$ of the vortex 
lattice and $H a_x a_y =\phi_0$, there is a relation 
\begin{equation} 
\Delta(\theta,{\bf r}+{\bf R}) =\Delta(\theta,{\bf r}) 
{\rm e}^{{\rm i}\chi({\bf r}, {\bf R})}, \quad 
{\bf a}({\bf r}+{\bf R})={\bf a}({\bf r}) , 
\end{equation}
where 
\begin{equation} 
\chi({\bf r}, {\bf R})=-\frac{\pi}{\phi_0} ({\bf H}\times{\bf R} )
\cdot({\bf r}+2{\bf r}_0)-\pi mn 
\end{equation}
in the symmetric gauge. 
Then, we can know $\Delta(\theta,{\bf r}) $ and ${\bf a}({\bf r})$ 
in the other region out of the calculated unit cell region. 
There is a vortex center at ${\bf r}_0-\frac{1}{2}({\bf r}_1+{\bf r}_2)$. 
When we consider the case when a vortex center locates at ${\bf r}=0$, 
we set ${\bf r}_0=\frac{1}{2}({\bf r}_1+{\bf r}_2)$. 
The spatial variation of the internal field and the current
${\bf J}({\bf r})=(c/4\pi)\nabla\times{\bf h}({\bf r})$ is calculated
from ${\bf a}({\bf r})$.
The current is scaled by $c \phi_0/4 \pi \xi_0$ in figures. 
To study the field dependence, our calculations are performed for various
fields at fixed temperature $T/T_c=0.5$.

In the chiral $p$-wave pairing, we can set 
\begin{eqnarray} 
\Delta(\theta,{\bf r})=\Delta_+({\bf r})\phi_+(\theta)+
 \Delta_-({\bf r})\phi_-(\theta), \\ 
V(\theta',\theta)=\bar{V}[ \phi_+^\ast(\theta')\phi_+(\theta)+
\phi_-^\ast(\theta')\phi_-(\theta) ] 
\end{eqnarray}
with the pairing functions $\phi_\pm (\theta)={\rm e}^{\pm{\rm i}\theta}$. 
For the $p_+$-wave domain case, we start our calculations 
from the initial state $\Delta_+({\bf r})=\Psi_0({\bf r})$ and 
$\Delta_-({\bf r})=0$ with the vortex lattice solution $\Psi_0({\bf r})$ 
in the lowest Landau level. In this case, $\Delta_-({\bf r})$ 
gives the induced component around the vortex after we obtain 
selfconsistent results. 
For the $p_-$-wave domain case, we start from the initial state 
$\Delta_-({\bf r})=\Psi_0({\bf r})$ and $\Delta_+({\bf r})=0$. 

When we consider the one component case for the pair potential by 
neglecting the induced component, 
$\Delta(\theta,{\bf r})=\Delta_+({\bf r})\phi_+(\theta)$ and  
$\Delta(\theta,{\bf r})=\Delta_-({\bf r})\phi_-(\theta)$ 
in the $p_+$-wave and the $p_-$-wave domain cases, respectively. 
In this case, by setting $\phi_\pm(\theta)
=|\phi_\pm(\theta)|{\rm e}^{\pm{\rm i}\theta}$, 
$f({\rm i}\omega_n,\theta,{\bf r}) 
=\bar{f}({\rm i}\omega_n,\theta,{\bf r}) {\rm e}^{\pm{\rm i}\theta}$ and 
$f^\dagger({\rm i}\omega_n,\theta,{\bf r})
={\bar f}^\dagger({\rm i}\omega_n,\theta,{\bf r}){\rm e}^{\mp{\rm i}\theta}$, 
we can remove the phase factor ${\rm e}^{\pm{\rm i}\theta}$ out of the 
Eilenberger equations. 
Then, the Eilenberger equations for 
$\bar{f}({\rm i}\omega_n,\theta,{\bf r})$ and 
$\bar{f}^\dagger({\rm i}\omega_n,\theta,{\bf r}) $ are solved under the 
pair potential $\Delta(\theta,{\bf r})
=\Delta_\pm({\bf r})|\phi_\pm(\theta)|=\Delta_\pm({\bf r})$, 
which is the pair potential for the isotropic $s$-wave pairing. 
Then, unless we consider the induced component, 
the vortex structure is the same as that of the $s$-wave pairing case, 
and there are no differences for the $p_+$-wave and the $p_-$-wave 
domain cases. 
Therefore, two component pair potential is intrinsic and essential 
for the vortex structure in the chiral $p$-wave superconductors. 
We also calculate the vortex structure in the isotropic $s$-wave pairing 
case for reference. 

The free energy $F$ is calculated as~\cite{Eilenberger,IchiokaQCLd2,KleinJLTP}
\begin{eqnarray}
\frac{F}{N_0 \Delta_0^2 }
&=&
 \kappa^2 \frac{\langle h({\bf r})^2 \rangle_{\bf r}}{(\phi_0/\xi_0^2)^2}
+ \frac{\langle |\Delta_+({\bf r})|^2 + |\Delta_-({\bf r})|^2 
     \rangle_{\bf r}}{N_0 \bar{V} \Delta_0^2}
\nonumber \\ &&
-\frac{2 \pi T}{\Delta_0^2}\sum_{\omega_n > 0}\int_0^{2\pi}
\frac{{\rm d} \theta}{2 \pi}\langle I(\omega_n,\theta,{\bf r}) \rangle_{\bf r}
\label{eq:free-energy}
\end{eqnarray}
with
\begin{eqnarray} &&
I(\omega_n,\theta,{\bf r})
=
\Delta^\ast(\theta,{\bf r}) f
+\Delta(\theta,{\bf r}) f^\dagger
\nonumber \\ &&
-( 1-g )
\Bigl[ \frac{1}{f}
\Bigl\{ \omega_n +{{\rm i} \over 2} {\bf v}_{\rm F}\cdot
\Bigl(\frac{\nabla}{\rm i}+\frac{2\pi}{\phi_0} {\bf A}({\bf r}) \Bigr)\Bigr\}f
\nonumber \\ &&
+\frac{1}{f^\dagger}
\Bigl\{ \omega_n -{{\rm i} \over 2} {\bf v}_{\rm F}\cdot
\Bigl(\frac{\nabla}{\rm i}-\frac{2\pi}{\phi_0} {\bf A}({\bf r}) \Bigr) \Bigr\}
f^\dagger\Bigr]
\label{eq:free-energy2} \\ &&
=\frac{\Delta^\ast(\theta,{\bf r}) f +\Delta(\theta,{\bf r}) f^\dagger}{1+g},
\label{eq:free-energy3}
\end{eqnarray}
where $g$, $f$ and $f^\dagger$ mean $g({\rm i}\omega_n,\theta,{\bf r}) $,
$f({\rm i}\omega_n,\theta,{\bf r}) $ and  
$f^\dagger({\rm i}\omega_n,\theta,{\bf r}) $, respectively.
For the spatial average, 
$\langle\cdots\rangle_{\bf r} 
=  \int_{\rm unit \ cell} {\rm d}{\bf r} (\cdots)/S$, 
where $S$ is the area of a unit cell. 
We use Eqs. (\ref{eq:eil-1})-(\ref{eq:eil-3}) to obtain
Eq. (\ref{eq:free-energy3}).

The LDOS for energy $E$ is calculated as
\begin{equation}
N(E,{\bf r})= N_0 \int_0^{2\pi} \frac{{\rm d}\theta}{2\pi}
{\rm Re}\ g({\rm i}\omega_n \rightarrow E+{\rm i}\eta,\theta,{\bf r}) .
\label{eq:ldos}
\end{equation}
To obtain $ g({\rm i}\omega_n \rightarrow E+{\rm i}\eta,\theta,{\bf r})$,
we solve Eqs. (\ref{eq:eil-1})-(\ref{eq:eil-3}) for $\eta-{\rm i}E$ instead
of $\omega_n$ using the self-consistently obtained $\Delta(\theta,{\bf r})$ 
and ${\bf a}({\bf r})$. 
We typically use $\eta=0.01$. 
The DOS is given by the spatial average of the LDOS as 
\begin{equation}
N(E)=\langle N(E,{\bf r}) \rangle_{\bf r}.
\label{eq:dos}
\end{equation}

\begin{figure} 
 \includegraphics[width=4cm]{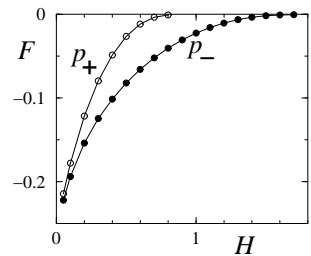} 
\caption{
Free energy as a function of $H$. 
We plot $F/(N_0 \Delta_0^2)$ for the $p_+$-wave domain case ($\circ$) 
and the $p_-$-wave domain case ($\bullet$) . 
} 
\label{fig:fe}
\end{figure} 

\section{Free energy} 
\label{sec:fenergy} 

For the estimation of the free energy difference between the $p_+$-wave 
domain and the $p_-$-wave domain cases, 
we show the field dependence of the free energy $F$ in 
Fig. \ref{fig:fe}.  
The $p_-$-wave domain case has lower free energy than that of the 
$p_+$-wave domain case. 
Then, the $p_-$-wave domain is stable, and the $p_+$-wave domain 
may exist as a metastable state. 
We expect that the transition of the $p_+$-wave domain to the 
$p_-$-wave domain is stimulated with increasing field, since the 
free energy difference increases. 
The upper critical field $H_{c2}\sim 0.81$ in the $p_+$-wave domain case, 
and  $H_{c2}\sim 1.7$ in the $p_-$-wave domain case. 
Then, the $p_+$-wave domain does not exist at $H>0.81$.  

The estimation of the stable vortex lattice configuration is also important, 
since the vortex lattice may be deformed from the triangular lattice 
due to the induced opposite chiral component.\cite{Scharnberg,ScharnbergL}  
Since our calculation method needs long computational time, 
we can not check all possible vortex lattice configurations. 
Then, we compare the free energy for the triangular and the square 
lattice configurations, and discuss the stable vortex lattice configuration. 
The free energy difference is less than $10^{-3}$ of $F$. 
We use finer $121 \times 121$ mesh within a unit cell 
to carefully estimate the difference. 
For higher field $H>0.35$, the square lattice configuration 
has lower free energy than the triangular one. 
It suggests that the square lattice is stable at higher field 
in the $p_-$-wave domain case. 
This is qualitatively consistent to the analysis by the GL theory, 
which suggests the square lattice configuration at higher 
field.\cite{Agterberg,Kita} 
And the square lattice is observed by neutron scattering 
experiment.\cite{Riseman} 
On the other hand, we obtain the result that 
the free energy of the triangular lattice is lower 
for all $H$ range in the $p_+$-wave domain case.  
It is because the induced opposite chiral component is small, as discussed 
later, and the vortex structure is similar to that of the isotropic 
$s$-wave pairing case. 
If the $p_-$-wave domain and the metastable $p_+$-wave domain coexist, 
we may observe the different vortex lattice configurations 
for the domains. 

In the following subsections, we investigate the origin of the difference 
of the vortex structure between the $p_-$-wave domain and the 
$p_+$-wave domains. 
The vortex structure is examined both in the square and the triangular 
vortex lattice configurations. 
Since our purpose is to find the chirality effect on the vortex structure 
by comparing the both chirality cases, we mainly report the results in 
the same situation of the square lattice configuration. 
The square lattice is expected in the stable $p_-$-wave domain case.   
After that, we briefly comment on the triangular lattice configuration case.

\begin{figure*} 
\includegraphics[width=16cm]{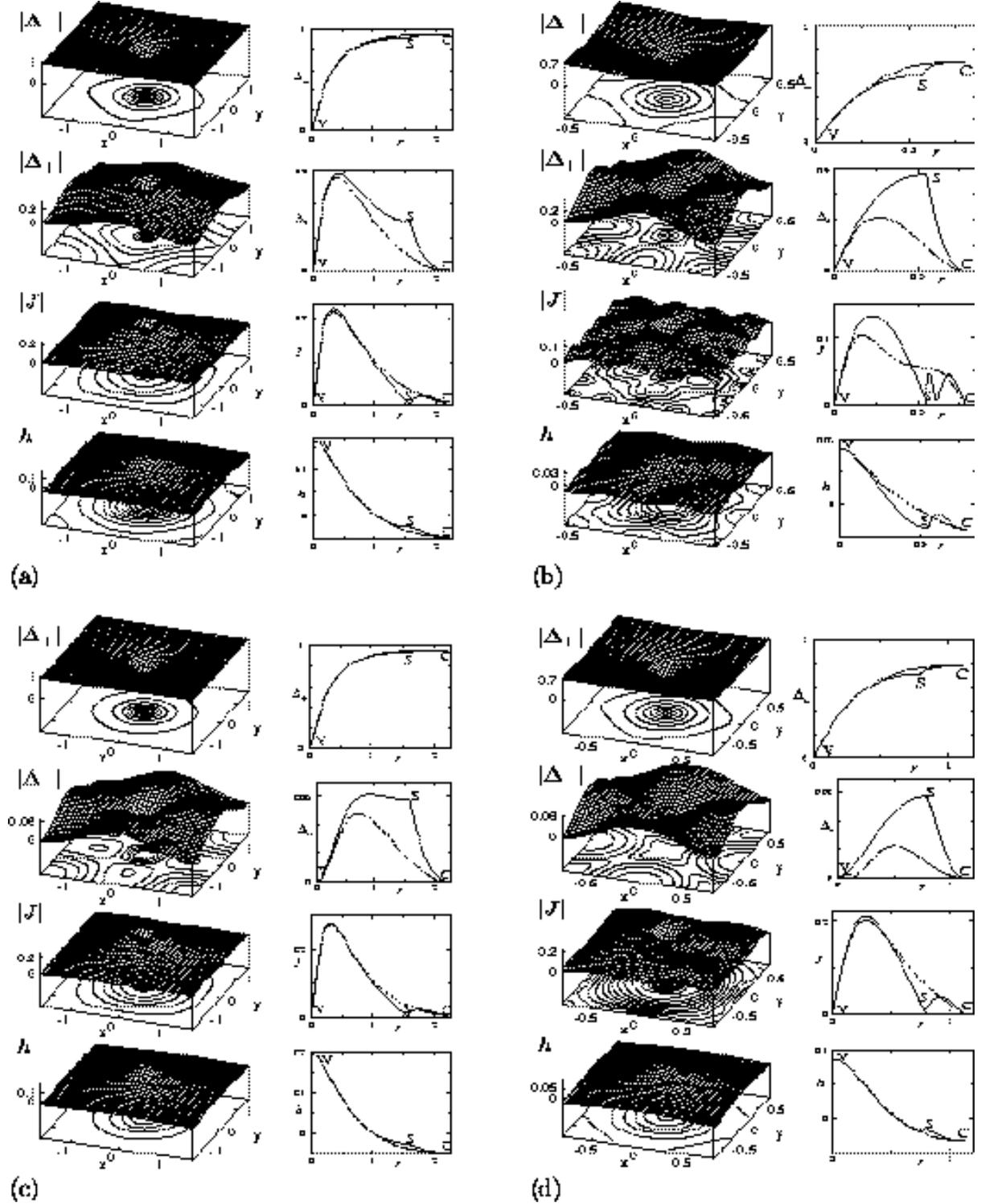}
\caption{
Vortex structure in the square vortex lattice configuration of the 
$p_-$-wave domain case at low field $H=0.1$ (a) and at high field 
$H=0.8$ (b), and in the $p_+$-wave domain case 
at $H=0.1$ (c) and $H=0.4$ (d).
From upper panels, we plot the dominant pair potential 
$|\Delta_\mp({\bf r})|$, the induced component $|\Delta_\pm({\bf r})|$, 
the screening current $|{\bf J}({\bf r})|$, and the internal field 
$h({\bf r})$, respectively. 
The left panels are the stereographic view within a unit cell region. 
There is a vortex at the center of the figures. 
The right panels are the profiles along the path V-S-C-V 
presented in Fig. \ref{fig:windS}(a). 
} 
\label{fig:dlt1}
\end{figure*} 
\begin{figure} 
\includegraphics{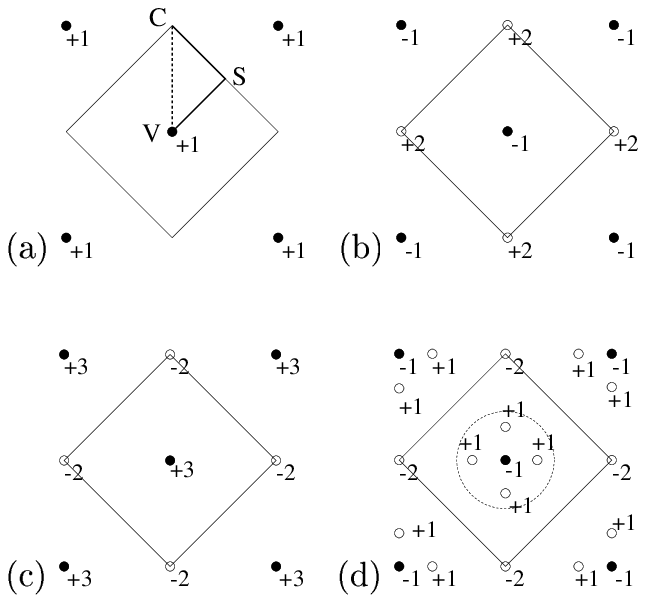}
\caption{
Phase winding structure of the pair potential 
in the square vortex lattice configuration. 
The vortex centers and other singularity points are, respectively, 
presented by $\bullet$ and $\circ$. 
We show the winding number for these points in the figures. 
The square area means a unit cell of the vortex lattice. 
(a) The phase of the dominant component. 
There is a winding $+1(\times 2\pi)$ at the vortex center. 
Along lines V-S-C-V, we consider profiles of the vortex structure 
in Fig. \ref{fig:dlt1}. 
(b) The phase of the induced pair potential $\Delta_+({\bf r})$ 
in the $p_-$-wave domain case. 
(c) The phase of the induced pair potential $\Delta_-({\bf r})$ 
at low field in the $p_+$-wave domain case. 
(d) The same as (c), but at high field. 
Total winding number is $+3$ within the dotted circle around the vortex core. 
}  
\label{fig:windS}
\end{figure} 

\section{Vortex structure}
\label{sec:vortex} 

\subsection{Vortex structure in the $p_-$-wave domain}

First, we consider the vortex structure in the $p_-$-wave domain case. 
It is shown in Fig. \ref{fig:dlt1} (a) and (b) 
within a unit cell of the square vortex lattice, i.e., square area 
in Fig. \ref{fig:windS} (a). 
The profiles are presented along the path V-S-C-V shown in 
Fig. \ref{fig:windS} (a). 
Line VS is along the nearest neighbor (NN) vortex direction, 
and line SC is along the boundary of the unit cell. 
Dashed line VC is along the next nearest neighbor (NNN) vortex direction. 
The dominant component $|\Delta_-({\bf r})|$ shows a 
conventional vortex structure.  
At low field, as shown in Fig. \ref{fig:dlt1}(a), 
$|\Delta_-({\bf r})|$  is recovered to $\Delta_0$ 
outside of the vortex core. 
The shape of the vortex core  is square-like. 
At higher field $H=0.8$ ($\sim 0.5 H_{c2}$) shown in Fig. \ref{fig:dlt1}(b), 
$|\Delta_-({\bf r})|$ is not recovered to $\Delta_0$ even in 
the boundary of the unit cell, since the inter-vortex distance is small. 
Along the NN vortex direction, $|\Delta_-({\bf r})|$ is slightly suppressed 
at the boundary region, compared to the NNN vortex direction.  

The opposite chiral $\Delta_+({\bf r})$ component is induced around the 
vortex core. 
At low field, the induced component is decayed outside of the vortex core. 
But at higher field, $|\Delta_+({\bf r})|$ has maximum at the S-points 
on the boundary.  
The induced component $\Delta_+({\bf r})$ always vanishes 
at the vortex center and at the C points. 
At the vortex center (the C points), $|\Delta_+({\bf r})|$ 
recovers with the $r$-liner ($r^2$-) dependence. 
These $r$-dependences are related to the phase winding of $\Delta_+({\bf r})$, 
which is presented in Fig. \ref{fig:windS} (b) schematically. 
At the vortex center, when the dominant component 
$\Delta_-({\bf r})$ has $+1(\times 2\pi)$ winding as shown in 
Fig. \ref{fig:windS} (a), the induced $\Delta_+({\bf r})$ component 
has opposite $-1$ winding. 
It is consistent with the results of previous 
theories.\cite{Heeb,HeebD,Kato,KatoHayashi,MatsumotoHeeb,Matsumoto,TakigawaP}  
Since the total of the winding should be $+1$ within a unit cell, 
$\Delta_+({\bf r})$ has also $+2$ winding at the C-points, i.e., 
corners of the square unit cell. 
These winding structures are the same for all $H$ range. 

The screening current $|{\bf J}({\bf r})|$ has maximum at the 
scale of the vortex core radius, and it is decreased with approaching 
the boundary of the unit cell. 
At the S- and C- points, $|{\bf J}({\bf r})|=0$. 
By this screening current, the internal field is produced. 
It has maximum at the vortex center, and it is decreased 
outside of the core. 
At low field, $h({\bf r})$ has minimum at C, and monotonically 
increases toward the S-point along the boundary. 
But, at high field, $|{\bf J}({\bf r})|$ and $h({\bf r})$ show 
anomalous behaviors at the boundary region. 
The profile of $h({\bf r})$ has a peak at a point between the 
S and the C points along the boundary line. 

\subsection{Vortex structure in the $p_+$-wave domain}

Next, we consider the vortex in the $p_+$-wave domain, which is shown 
in Fig. \ref{fig:dlt1}(c) and (d).  
The vortex core shape of the dominant component 
shows circular shape at low field, as shown in the contour line of 
$|\Delta_+({\bf r})|$ in Fig. \ref{fig:dlt1} (c). 
The induced component $\Delta_-({\bf r})$ becomes zero at the vortex center 
and the C-points also in this case. 
The amplitude $|\Delta_-({\bf r})|$ recovers with the $r^2$-behavior around 
the corners C, as in the $p_-$-wave domain case. 
But, the recovery at the vortex center shows the $r^3$-behavior, 
instead of the $r$-linear. 
It is because $\Delta_-({\bf r})$ has $+3$ winding at the vortex center, 
as schematically presented in Fig. \ref{fig:windS} (c). 
The winding $-2$ at the C-point is not changed. 
Compared with the $p_-$-wave domain case, internal field $h({\bf r})$ 
at the vortex core is larger, since  $|{\bf J}({\bf r})|$ around the 
vortex is larger. 

At high field, the winding structure of the induced component 
$\Delta_-({\bf r})$ is changed around the vortex center.
The $+3$ winding at the vortex center splits to a $-1$ winding at the 
center and four $+1$ winding points around the core, 
as schematically shown in Fig. \ref{fig:windS} (d). 
As is seen in Fig. \ref{fig:dlt1}(d), $|\Delta_-({\bf r})|=0$ 
at these  $+1$ winding points. 
As for $|{\bf J}({\bf r})|$ and $h({\bf r})$, 
the high field case shows the similar structure as in the low field 
case, while their strengths are suppressed with increasing field. 

\begin{figure} 
\includegraphics{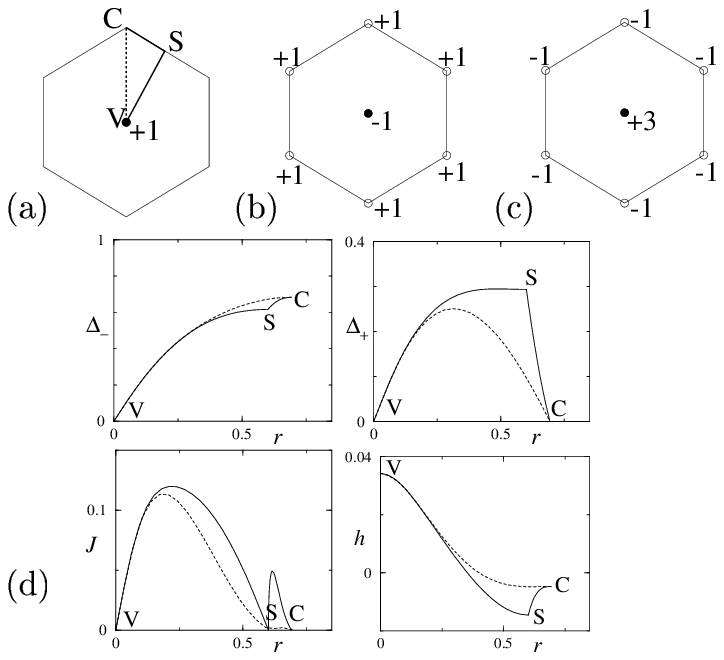}
\caption{
Phase winding structure in the triangular vortex lattice configuration 
for the dominant component (a), and the induced component 
in the $p_-$-wave domain case (b) and the $p_+$-wave domain case (c). 
The vortex centers and other singularity points are, respectively, 
presented by $\bullet$ and $\circ$. 
We show the winding number for these points in the figures. 
The hexagonal area means a unit cell of the vortex lattice. 
(d) The profile of the vortex structure in the triangular vortex 
lattice configuration of the $p_-$-wave domain case at high field 
$H=0.8$. 
We plot the dominant pair potential $|\Delta_-({\bf r})|$, 
the induced component $|\Delta_+({\bf r})|$, 
the screening current $|{\bf J}({\bf r})|$ and 
the internal field $h({\bf r})$ 
along the path V-S-C-V presented in (a). 
}  
\label{fig:windT}
\end{figure} 

\subsection{In the triangular lattice configuration}

We briefly report the vortex structure in the triangular vortex lattice 
configuration. 
In this case, the winding structure at the corner of the unit cell 
is changed. 
In the $p_-$-wave ($p_+$-wave) domain case, $+2$ ($-2$) winding 
in the square lattice becomes $+1$ ($-1$) winding at the 
corners of the hexagonal unit cell in the triangular lattice, 
as shown in Fig. \ref{fig:windT} (a)-(c). 
In the square lattice case, the winding structure at the vortex center 
changes from $+3$ to $-1$ on raising field in the $p_+$-wave domain case
[Fig. \ref{fig:windS} (c) and (d)], 
the vortex center keeps $+3$ winding for all $H$ range 
in the triangular lattice configuration. 

As an example, we show the profile of the vortex structure for 
the $p_-$-wave domain at $H=0.8$ in Fig. \ref{fig:windT}(d), 
which is plotted along the NN direction (VS line in Fig. \ref{fig:windT}(a)), 
boundary line (SC line) and the NNN direction (VC line) 
in the triangular vortex lattice. 
Compared with Fig. \ref{fig:dlt1}(d) in the square lattice configuration, 
the recovery of the induced component around the C-points shows 
$r$-linear relation instead of the $r^2$-behavior, 
reflecting the change of the winding structure. 
There appears anomalous field distribution at higher field 
in the $p_-$-wave domain. 
But, its profile is different from that of the square lattice case. 
In the triangular lattice, $h({\bf r})$ has minimum at the S-points. 
However, the $p_+$-wave domain case and the low field $p_-$-wave domain case  
show qualitatively the same field distribution as that of the square 
lattice case in the profile plot. 
There, $h({\bf r})$ has minimum at the C points. 

\begin{figure} 
\includegraphics{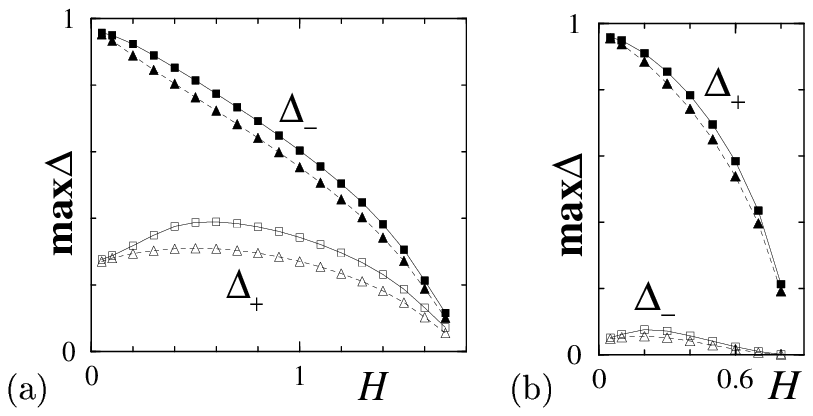}
\caption{
Field dependence of the amplitude of the 
dominant and the induced pair potentials in the $p_-$-wave domain 
case (a) and in the $p_+$-wave domain case (b). 
We plot ${\rm max}|\Delta_-({\bf r})|$ and 
${\rm max}|\Delta_+({\bf r})|$ as a function of $H$. 
Solid (dashed) lines are for the square (triangular) vortex lattice 
configuration. 
}  
\label{fig:maxd}
\end{figure} 
\begin{figure} 
\includegraphics{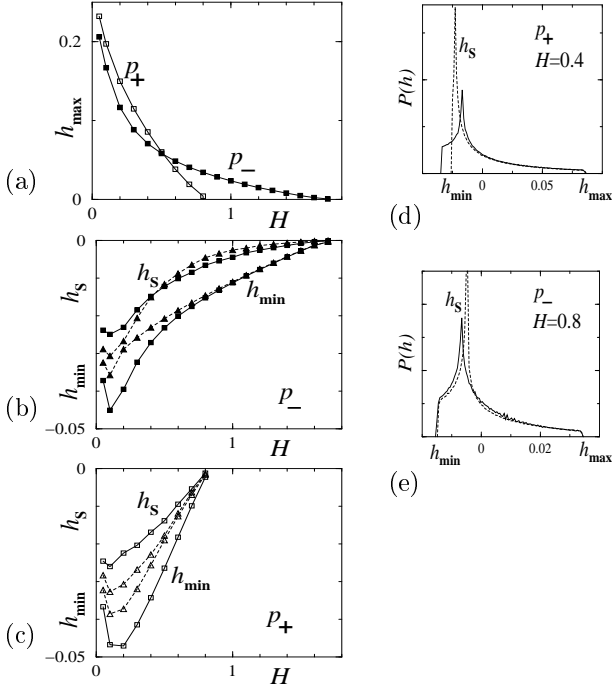}
\caption{
Field dependence of the internal field distribution. 
(a) Internal field at the vortex center, $h_{\rm max}$, 
in the $p_-$-wave and the $p_+$-wave domain cases. 
Both the square and the triangular lattice configurations give  
same $h_{\rm max}$. 
(b) The minimum $h_{\rm min}$ of the internal field, and the 
peak field $h_{\rm s}$ of the distribution function $P(h)$ in the 
$p_-$-wave domain case. 
Solid (dashed) lines are for the square (triangular) vortex lattice 
configuration. 
(c) The same as (b), but in the $p_+$-wave domain case.  
(d) Internal field distribution function $P(h)$ at 
$H=0.4$ in the $p_+$-wave domain case. 
Solid (dashed) lines are for the square (triangular) vortex lattice 
configuration.  
(e)  The same as (d), but at $H=0.8$ in the $p_-$-wave domain case. 
}  
\label{fig:hd}
\end{figure} 

\subsection{Magnetic field dependence}

In this subsection, we investigate the continuous field dependence 
of the vortex structure. 
The field dependence of $\Delta_-^{\rm max}={\rm max}|\Delta_-({\bf r})|$ 
and  $\Delta_+^{\rm max}={\rm max}|\Delta_+({\bf r})|$ is presented 
in Fig. \ref{fig:maxd}. 
We show it both for the square lattice and the triangular lattice 
configurations. 
In the $p_-$-wave domain case, the induced $\Delta_+^{\rm max}$ 
is large. 
When the amplitude of the dominant $\Delta_-^{\rm max}$ is decreased 
with raising field, the ratio of the induced component, 
$\Delta_+^{\rm max}/\Delta_-^{\rm max}$, increases monotonically 
up to $H_{c2}$. 
Due to this large induced component, the superconductivity in the $p_-$-wave 
domain case can survive until higher magnetic field, giving high $H_{c2}$. 

On the other hand, in the $p_+$-wave domain case, the induced component 
$\Delta_-^{\rm max}$ is small. 
The ratio of the induced component, $\Delta_-^{\rm max}/\Delta_+^{\rm max}$, 
decreases as a function of $H$ at higher field, after increasing at low field. 
Since the ratio is reduced to zero at  $H \rightarrow H_{c2}$, 
the $p_+$-wave domain has the same $H_{c2}$ as in the isotropic  
$s$-wave pairing in the two dimensional Fermi surface case.  
We also calculate the isotropic $s$-wave case, 
which is equivalent to the case when we neglect the induced 
component of the chiral $p$-wave case. 
The field dependence of ${\rm max}|\Delta({\bf r})|$ in the $s$-wave 
pairing is almost the same as that of $\Delta_+^{\rm max}$ in 
Fig. \ref{fig:maxd} (b). 
The large amplitude of the induced component in the $p_-$-wave 
domain is the origin of the different behavior of the vortex structure 
between the $p_+$-wave domain and the $p_-$-wave domain cases. 

The magnetic field dependence of the internal field distribution 
is shown in Fig. \ref{fig:hd}. 
There, we plot $H$-dependence of the maximum $h_{\rm max}$ and the minimum 
$h_{\rm min}$ of $h({\bf r})$. 
The $p_+$-wave case and the $s$-wave pairing case have similar 
internal field distributions. 
$h_{\rm max}$ is the internal field at the vortex center. 
Compared with the $p_-$-wave domain case, 
$h_{\rm max}$ in the $p_+$-wave domain case is larger at low $H$, and 
becomes smaller at higher $H$, because $h_{\rm max}\rightarrow 0$ with 
approaching low $H_{c2}$. 
We obtain same  $h_{\rm max}$ both in the square and the triangular 
vortex lattice configurations. 

The internal field distribution function is defined as 
\begin{equation} 
P(h)=\langle \delta( h - h({\bf r}) )\rangle_{\bf r} 
\end{equation} 
with the $\delta$-function. 
We can observe $P(h)$ as the resonance line shape in the NMR or 
$\mu$SR experiments. 
Figure \ref{fig:hd}(d) shows $P(h)$ at $H=0.4$ in the $p_+$-wave domain case. 
It is a typical distribution shape as in the conventional superconductor, 
i.e., $h_{\rm min}$ is the internal field at the C-point on the unit cell 
boundary, and peak field $h_{\rm s}$ corresponds to the internal field at the 
saddle point S.\cite{Fetter} 
The vortex lattice configuration affects the distance between $h_{\rm s}$ and 
$h_{\rm min}$. 
Compared with the square lattice configuration, $h_{\rm s}-h_{\rm min}$ 
is small in the triangular lattice configuration. 
Figure \ref{fig:hd}(c) shows that this property of $P(h)$ appears 
for all $H$ region in the $p_+$-wave domain case.  

Figure \ref{fig:hd}(e) shows  $P(h)$ at $H=0.8$ in the $p_-$-wave domain case, 
which gives anomalous distribution of $h({\bf r})$ at the unit cell boundary, 
as shown in Figs. \ref{fig:dlt1}(b) and \ref{fig:windT}(d).  
There, the position giving $h_{\rm min}$ ($h_{\rm s}$) shifts to 
other position than the C- (S-) point on the boundary. 
In this distribution, the square and the triangular lattice configurations 
give similar $P(h)$ distribution. 
When we see the field dependence of $h_{\rm s}$ and $h_{\rm min}$ in 
Fig. \ref{fig:hd}(b), there is a small difference between the triangular 
and the square lattice configurations at higher field $H > 0.4$. 
But, at lower field $H < 0.4$, there appears eminent difference 
in the distance between $h_{\rm s}$ and $h_{\rm min}$ for 
the two vortex lattice configurations. 
There $P(h)$ has similar distribution as in Fig. \ref{fig:hd}(d). 
It is interesting that the square vortex lattice configuration has 
lower free energy than the triangular one in the field region 
$H > 0.4$, showing anomalous field distribution. 

\begin{figure} 
\includegraphics[width=8cm]{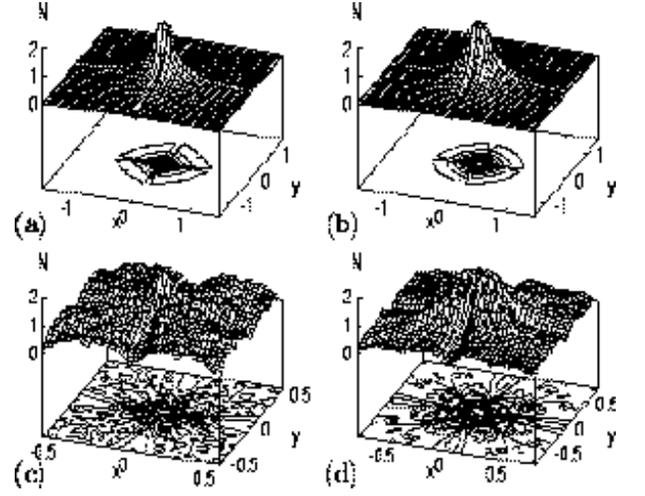}
\caption{
Zero-energy local density of states in the $p_-$-wave domain case 
at $H=$0.1 (a), 0.8 (c), and  in the $p_+$-wave domain case 
at $H=$0.1 (b), 0.4 (d). 
We plot $N(E=0,{\bf r})/N_0$ within a unit cell. 
The peak at the vortex center is truncated at $N(E=0,{\bf r})/N_0=2$ 
to show the tail structure extending from the vortex core.  
}  
\label{fig:LDOS}
\end{figure} 
\begin{figure} 
\includegraphics{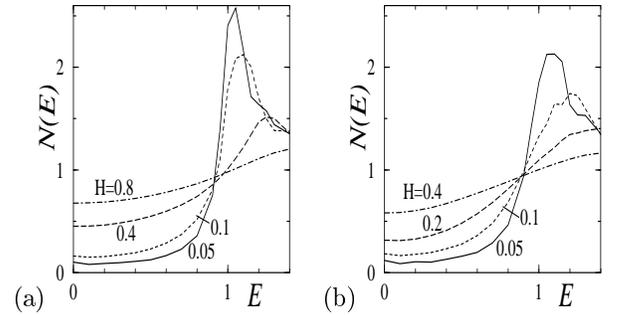}
\caption{
Spectrum of the spatially averaged DOS, $N(E)/N_0$. 
(a) In the $p_-$-wave domain case at $H=$0.05, 0.1, 0.4, 0.8. 
(b) In the $p_+$-wave domain case at $H=$0.05, 0.1, 0.2, 0.4.  
The square and the triangular vortex lattice configurations give 
same result. 
The spectrum of the $p_+$-wave domain case  is almost the same as 
in the $s$-wave pairing case. 
}  
\label{fig:spectrum}
\end{figure} 
\begin{figure} 
\includegraphics{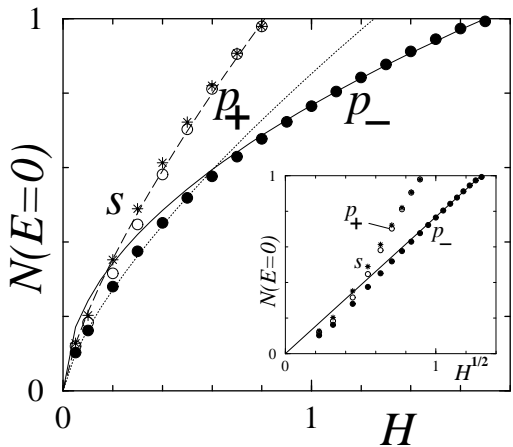}
\caption{
Field dependence of the zero-energy DOS $N(0)/N_0$ 
in the $p_-$-wave domain case ($\bullet$), the $p_+$-wave 
domain case ($\circ$), and the $s$-wave case ($\ast$). 
Points show numerical data, and lines are fitting curves by 
$N(0)\propto H^\alpha$. 
Dashed line and dotted line are for $\alpha=$0.74 and 0.71, respectively. 
Solid line is for $N(0)\propto \sqrt{H}$. 
Inset: $N(0)/N_0$ is plotted as a function of $\sqrt{H}$. 
The solid line shows the relation $N(0)\propto \sqrt{H}$. 
}  
\label{fig:DOS}
\end{figure} 
\begin{figure} 
\includegraphics[width=8cm]{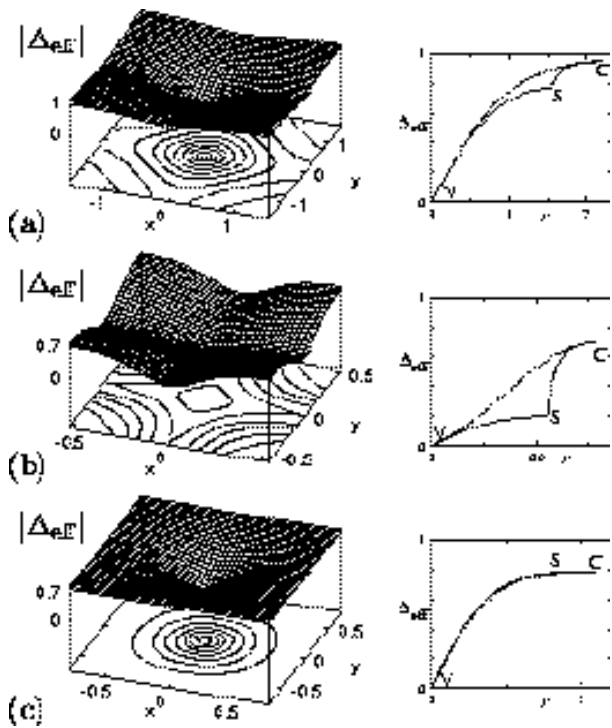}
\caption{
Effective pair potential for zero-energy quasiparticles. 
We plot $|\Delta_{\rm eff}({\bf r})|=|\Delta(\theta_{\bf r},{\bf r})|$ 
at low field $H=0.1$ (a) and at higher field $H=0.8$ (b) 
in the $p_-$-wave domain case, and at $H=0.4$ in the $p_+$-wave 
domain case (c).  
The left panels are the stereographic view within a unit cell region. 
There is a vortex at the center of the figures. 
The right panels are the profiles along the path presented in Fig. 
\ref{fig:windS}(a). 
}  
\label{fig:dlteff}
\end{figure} 

\section{Quasiparticle structure in the vortex state}
\label{sec:LDOS}

The LDOS is expected to be observed by the scanning tunneling microscopy 
(STM), which will experimentally give the detailed information of the 
quasiparticle structure around vortices. 
We study the LDOS at zero energy, which dominantly contributes 
to the low temperature behaviors. 
Figures \ref{fig:LDOS} (a) and (b) show $N(E=0,{\bf r})$ at low field 
$H=0.1$ in the $p_-$-wave and the $p_+$-wave domain cases, respectively. 
Since the gap function has full gap as in the $s$-wave pairing, 
the low energy states are localized at the vortex core. 
We see that the localized LDOS is slightly suppressed along the 
NN and NNN vortex directions. 
It is the effect of the inter-vortex transfer of the low energy 
quasiparticles.~\cite{IchiokaQCLd1,IchiokaQCLd2,IchiokaJS,IchiokaQCLs,KleinPRB} 

Higher field case at $H \sim 0.5 H_{c2}$ are presented in Fig. 
\ref{fig:LDOS} (c) and (d). 
Figure \ref{fig:LDOS} (c) is for the $p_-$-wave domain case at $H=0.8$, 
and Fig. \ref{fig:LDOS} (d) is for the $p_+$-wave domain case at $H=0.4$. 
Since the inter-vortex distance becomes short at high field, 
the LDOS localized at vortex cores are overlapped each other. 
We see the eminent suppression along the NN and NNN vortex directions 
due to the inter-vortex transfer. 
The localized LDOS around the vortex core is reduced to uniform 
distribution when approaching $H_{c2}$. 
Then, the sharp peak at the vortex center survives until higher field 
in the $p_-$-domain case, since it has higher $H_{c2}$. 

The spectrum of the spatially averaged DOS, $N(E)$, is presented in 
Fig. \ref{fig:spectrum}. 
The full gap structure for $E < \Delta_0$ and the peak of the gap edge  
at $E = \Delta_0$ are gradually smeared by low energy quasiparticles 
around the vortex. 
These behaviors of the LDOS and spectrum $N(E)$ are the same as 
in the isotropic $s$-wave case previously 
reported.\cite{IchiokaQCLd1,IchiokaQCLd2,IchiokaJS,IchiokaQCLs}  
When we see the spatial structure of the LDOS, we do not find 
drastic changes by the chirality effect. 
We see qualitatively the same structure both for the $p_-$-wave 
domain and $p_+$-wave domain cases. 

However, when we consider the quantitative field dependence of 
the DOS $N(0)$ by spatially averaging the LDOS at $E=0$, 
we can see the characteristic behavior 
of the chiral $p$-wave pairing. 
Figure \ref{fig:DOS} shows the field dependence of $N(0)$. 
Numerical data are presented by points, and lines are fitting 
curves by the relation $N(0)\propto H^\alpha$. 
The square and the triangular vortex lattice configurations give 
the same result. 

In the $p_+$-wave domain case, $N(0)$ shows similar behavior 
as that of the $s$-wave pairing case. 
But $N(0)$ is slightly smaller than that of the $s$-wave case at low field. 
Fitting curve is given by $\alpha=0.74$ in the $s$-wave case for 
$\kappa=2.7$. 
In the $p_-$-wave domain case, lower field data are fitted by $\alpha=0.71$. 
But, higher field data are fitted by $\alpha=0.5$. 
To see the $\sqrt{H}$-behavior, we plot $N(0)$ as a function of $\sqrt{H}$ 
in the inset. 
At higher field, $N(0)$ is on the line $N(0)\propto \sqrt{H}$. 
This $\sqrt{H}$-behavior at higher field is qualitatively consistent 
with the experimental data of the specific heat.\cite{Deguchi}  
We note that this $\sqrt{H}$ is not the so-called Volovik effect   
for the vertical line node of the 
$k_x^2-k_y^2$-type.\cite{IchiokaQCLd1,Volovik} 
In the vertical line node case, the $\sqrt{H}$-behavior appears  
from low field. 
According to the recent directional dependent thermal-conductivity 
experiment under parallel field,\cite{Izawa} 
there is no in-plane gap anisotropy, implying the 
absence of the vertical line node. 

Unless we consider the induced component of the pair potential, 
$|\Delta(\theta,{\bf r})|=|\Delta_-({\bf r})||\phi_-(\theta)|$ 
gives full gap, as in the $s$-wave pairing case. 
Then, we can not expect the $\sqrt{H}$-behavior of $N(0)$. 
However, when we take account of the induced component, 
$|\Delta(\theta,{\bf r})|=|\Delta_-({\bf r})\phi_-(\theta)
+\Delta_+({\bf r})\phi_+(\theta)|$ can be small for particular 
direction $\theta$, if the induced component $|\Delta_+({\bf r})|$ is large. 
To discuss the origin of the $\sqrt{H}$-behavior at higher field, 
we consider the effective pair potential for zero-energy quasiparticles. 
When we analyze the LDOS at $E=0$, the main contribution comes from the 
quasiparticle trajectory passing through the vortex center (i.e., line with 
the impact parameter $r_\perp =0$),\cite{KleinPRB} since zero-energy 
quasiparticles are localized around the vortex core. 
Then, the zero-energy LDOS at ${\bf r}=(x,y)$ dominantly consists of 
quasiparticles traveling along the quasiparticle trajectory with the 
direction $\theta_{\bf r}=\tan^{-1}(y/x)$. 
They feel the effective pair potential 
$\Delta_{\rm eff}({\bf r})\equiv \Delta(\theta_{\bf r},{\bf r})
=\Delta_+({\bf r}) \phi_+(\theta_{\bf r}) 
+\Delta_-({\bf r}) \phi_-(\theta_{\bf r})$.  

Figure \ref{fig:dlteff} shows amplitude $|\Delta_{\rm eff}({\bf r})|$ 
for some typical cases. 
In the $p_-$-wave domain case, 
$|\Delta_{\rm eff}({\bf r})|\sim |\Delta_-({\bf r})|-|\Delta_+({\bf r})|$, 
i.e., the induced component $|\Delta_+({\bf r})|$ suppresses 
the effective pair potential. 
At low field, since the induced component is localized around the vortex core, 
$|\Delta_{\rm eff}({\bf r})|$ is suppressed around the vortex core 
as shown in Fig. \ref{fig:dlteff}(a). 
Since the amplitude of the effective pair potential is still large 
at the boundary of the unit cell, we expect the bound quasiparticle 
states around the vortex core. 
Then, the exponent $\alpha \sim 0.71 $ is not largely different from that of  
the $s$-wave case. 
At higher $H$, the ratio of the induced component is enhanced, 
and $|\Delta_+({\bf r})|$ has large 
amplitude at the boundary of the unit cell. 
Since the induced component $|\Delta_+({\bf r})|$ is large 
in the NN vortex direction [Fig. \ref{fig:dlt1}(b)], 
$|\Delta_{\rm eff}({\bf r})|$ is largely suppressed in this direction, 
as shown by line VS in Fig. \ref{fig:dlteff}(b). 
The effective pair potential is not suppressed at the C-points, 
since there is no induced component there. 
Also in the triangular vortex lattice configuration, 
$|\Delta_{\rm eff}({\bf r})|$ gives the similar structure. 
There, $|\Delta_{\rm eff}({\bf r})|$ is eminently suppressed along six 
NN vortex directions. 
With increasing $H$, $|\Delta_{\rm eff}({\bf r})|$ along the NN direction 
is more suppressed, since the ratio  
$\Delta_+^{\rm max}/\Delta_-^{\rm max}$ increases monotonically. 
Then, the low energy quasiparticles can easily transfer between NN vortices 
at higher field.  
When low energy quasiparticles are extended to the boundary of the unit 
cell as in the $d$-wave pairing case, we expect the $\sqrt{H}$-like 
behavior.\cite{Volovik,IchiokaQCLd1} 
It is the origin of the relation $N(0)\sim\sqrt{H}$ at high field.  

On the other hand, in the $p_+$-wave domain case, 
$|\Delta_{\rm eff}({\bf r})|\sim |\Delta_+({\bf r})|+|\Delta_-({\bf r})|$, 
i.e., the induced component $|\Delta_-({\bf r})|$ enhances 
the effective pair potential, as shown in Fig. \ref{fig:dlteff}(c).  
Then, low energy quasiparticles are bound states, 
as in the $s$-wave pairing case in all $H$ range. 
Due to the enhancement of the effective pair potential by the induced 
component, $N(0)$ is slightly suppressed than that of the $s$-wave case. 
With approaching $H_{c2}$, 
$N(0)$ is reduced the $s$-wave case's value, 
since the induced component is decreased to zero. 

\section{Summary and discussions}
\label{sec:summary}

We have investigated the field dependence of the vortex structure 
in chiral $p$-wave superconductors. 
We have shown the difference of the vortex structure for 
the $p_+$-wave domain case and the $p_-$-wave domain case. 
The difference comes from the structure of the induced 
opposite chiral component. 
When we compare the free energy, the $p_-$-wave domain is the stable 
state, and the $p_+$-wave domain is metastable. 
Then, the transition of the $p_+$-wave domain to the $p_-$-wave domain 
may occur. 
We expect different vortex lattice configuration for the domains, i.e., 
square-like lattice in the $p_-$-wave domain and triangular lattice 
in the $p_+$-wave domain. 

The phase winding structure of the induced component of the pair potential 
is  different depending on the chirality. 
In the $p_+$-wave domain, the amplitude of the induced component 
is small and reduced to zero near $H_{c2}$. 
Then, the vortex structure is similar to that of the 
isotropic $s$-wave pairing case. 
And $H_{c2}$ is same as in the $s$-wave pairing case 
in the two-dimensional Fermi surface. 
In the $p_-$-wave domain case, the opposite chiral component 
is largely induced. 
Then, the superconductivity can survive until high field, giving 
high $H_{c2}$.  
The induced component produces the characteristic vortex structure 
in the chiral $p$-wave superconductors, such as an anomalous 
internal field distribution. 

The LDOS structure shows that low energy quasiparticles are bound 
states around the vortex core, and there are some inter-vortex transfers. 
At higher field, the bound states are overlapped between neighbor vortices. 
When we quantitatively consider the field dependence of the 
zero energy DOS $N(0)\propto H^\alpha$, we obtain the effect of the 
chiral $p$-wave superconductivity. 
The stable $p_-$-wave domain case shows $\sqrt{H}$-behavior 
at higher field. 
It is because the effective pair potential for zero energy quasiparticles 
are suppressed along the NN vortex direction by the induced opposite 
chiral component of the pair potential. 
At low field, the suppression by the induced component is restricted 
in the vortex core region, 
the low energy quasiparticles are still bound around the vortex core. 
Then the exponent $\alpha$ is near the value for the $s$-wave pairing 
at low field. 

The superconducting state in ${\rm Sr_2 Ru O_4}$ is suggested to be the 
chiral $p$-wave pairing. 
If we can experimentally observe the domain structure of the $p_+$-wave 
and the $p_-$-wave pairing regions, it becomes firm evidence 
for the chiral $p$-wave superconductivity. 
In this observation, the information of the vortex structure difference 
for the $p_\pm$-wave domains is helpful to analyze the chirality 
$p_\pm$ of each domain. 
For example, the internal field distribution and the stable vortex lattice 
configuration may be different depending on the chirality of the domain. 
The specific heat measurement on ${\rm Sr_2 Ru O_4}$ reports that 
$N(0)\propto \sqrt{H}$ at high field, while it deviates from $\sqrt{H}$ 
at low field.\cite{Deguchi}  
It is qualitatively consistent with our results. 
However, when we analyze the experimental data on ${\rm Sr_2 Ru O_4}$, 
there are some factors to quantitatively modify our results 
on a simple isotropic system, such as the possibility of the line node 
along the basal plane direction, the orbital dependence and anisotropy 
of the Fermi surface and the gap 
functions.~\cite{Machida,Sigrist,Hasegawa,Graf,Zhitomirsky,Ng,Miyake} 
The study on these additional effects remains in future problems.

\begin{acknowledgments}

We would like to thank N. Hayashi, M. Takigawa, N. Nakai, P. Miranovic and 
M. Sigrist for their helpful comments and discussions. 

\end{acknowledgments}
\appendix
\section{Symmetry relation}

When one of the vortex center locates at ${\bf r}=0$, there is a relation 
$\Delta(\theta,{\bf r})=-\Delta(\theta,-{\bf r})$. 
Then, considering the transformation ${\bf r}\rightarrow -{\bf r}$ in 
the Eilenberger equations (\ref{eq:eil-1})-(\ref{eq:eil-3}), 
we obtain the following relations of the quasiclassical Green's functions, 
\begin{eqnarray} &&
f(\omega_n,\theta,-{\bf r})=-f^{\dagger\ast}(\omega_n^\ast,\theta,{\bf r}), 
\nonumber \\ &&
f^\dagger(\omega_n,\theta,-{\bf r})=-f^\ast(\omega_n^\ast,\theta,{\bf r}), 
\nonumber \\ &&
g(\omega_n,\theta,-{\bf r})=g^\ast(\omega_n^\ast,\theta,{\bf r}). 
\end{eqnarray} 
Then, in the calculation of the Matsubara frequency $\omega_n$ or $E=0$, 
it is enough to solve the Eilenberger equations 
in half area of a unit cell. 

When the vortex lattice is symmetric under the reflection at the $x$ axis,  
i.e., $S{\bf r}=(x,-y)$, there is a relation 
$\Delta(-\theta,S{\bf r})
=-{\rm e}^{{\rm i}\alpha }\Delta^\ast(\theta,{\bf r})$. 
The factor ${\rm e}^{{\rm i}\alpha }$ comes from 
$\phi(-\theta)={\rm e}^{{\rm i}\alpha }\phi^\ast(\theta)$ 
for the pairing function of the dominant component of the pair potential. 
Then, considering the transformation $S{\bf r}$ and 
$\theta \rightarrow -\theta$ in Eqs. (\ref{eq:eil-1})-(\ref{eq:eil-3}), 
we obtain the following relations of the quasiclassical Green's functions, 
\begin{eqnarray} &&
f(\omega_n,-\theta,S{\bf r})=-{\rm e}^{{\rm i}\alpha }
f^\ast(\omega_n,\theta,{\bf r}) , 
\nonumber \\ &&
f^\dagger(\omega_n,-\theta,S{\bf r})=-{\rm e}^{-{\rm i}\alpha }
f^{\dagger\ast}(\omega_n,\theta,{\bf r}), 
\nonumber \\ &&
g(\omega_n,-\theta,S{\bf r})=g^\ast(\omega_n,\theta,{\bf r}). 
\end{eqnarray} 

Next, we consider the $\psi$-rotation around the vortex center 
at ${\bf r}=0$, i.e., $R_\psi {\bf r}=(x \cos\psi -y \sin\psi, 
x \sin\psi +y \cos\psi)$.  
Generally, vortex lattice has the symmetry for the rotation $\psi=\pi$. 
And further, the square (triangular) vortex lattice has the 
symmetry for the rotation $\psi=\pi/2$ ($\psi=\pi/3$).
Under these rotations, 
$\Delta(\theta+\psi,R_\psi{\bf r})={\rm e}^{{\rm i}(\alpha' -\psi)} 
\Delta(\theta,{\bf r})$ in the symmetric gauge. 
The factor ${\rm e}^{{\rm i}\alpha' }$ comes from 
$\phi(\theta+\psi)={\rm e}^{{\rm i}\alpha' }\phi(\theta)$. 
Then, considering the translation $R_\psi {\bf r}$ and 
$\theta \rightarrow \theta +\psi$ in Eqs. (\ref{eq:eil-1})-(\ref{eq:eil-3}), 
we obtain 
\begin{eqnarray} &&
f(\omega_n,\theta+\psi,R_\psi{\bf r})
={\rm e}^{{\rm i}(\alpha' -\psi)}f(\omega_n,\theta,{\bf r}), 
\nonumber \\ &&
f^\dagger(\omega_n,\theta+\psi,R_\psi{\bf r})
={\rm e}^{-{\rm i}(\alpha' -\psi)}f^\dagger(\omega_n,\theta,{\bf r}), 
\nonumber \\ &&
g(\omega_n,\theta+\psi,R_\psi {\bf r})=g(\omega_n,\theta,{\bf r}). 
\end{eqnarray} 
In the pair potential $\Delta(\theta,{\bf r})
=\Delta_-({\bf r})\phi_-(\theta)+\Delta_+({\bf r})\phi_+(\theta)$, 
the pairing function $\phi_\pm(\theta)$ of the induced component 
may produce different phase factor from that of 
the dominant component  $\phi_\mp(\theta)$ in the transformation 
$\phi(\theta+\psi)={\rm e}^{{\rm i}\alpha' }\phi(\theta)$. 
Then, the phase of the induced component $\Delta_\pm({\bf r})$ should 
be changed so as to cancel the difference of the phase factors 
in the rotational transformation.  
This is the origin of the different phase winding of the induced 
component in Figs. \ref{fig:windS} and \ref{fig:windT}. 

Using these symmetry relations, it is enough to solve 
the Eilenberger equations (\ref{eq:eil-1})-(\ref{eq:eil-3}) 
for $0 \le \theta < \pi/4$ ($0 \le \theta < \pi/6$) 
in the square (triangular) vortex lattice configuration.


\end{document}